\documentclass[journal,twocolumn]{IEEEtran}
\usepackage{amsfonts}
\usepackage[tbtags]{amsmath}
\usepackage{graphicx,subfigure}
\usepackage{multirow}
\usepackage{color}      
\usepackage{epsfig}
\usepackage{amssymb}
\usepackage{tipa}
\usepackage{bbm}

\def\endproof{\hspace*{\fill}~$\blacksquare$}
\long\def\comment#1{}

\newcommand{\beq}{\begin{equation}}
\newcommand{\eeq}{\end{equation}}
\newcommand{\beqno}{\begin{equation*}}
\newcommand{\eeqno}{\end{equation*}}

\newcommand{\bes}{\begin{split}}
\newcommand{\ees}{\end{split}}
\newcommand{\bdm}{\begin{displaymath}}
\newcommand{\edm}{\end{displaymath}}

\newtheorem{definition}{Definition}

\newcommand{\bd}{\begin{definition}}
\newcommand{\ed}{\end{definition}}

\newcommand{\bfx} {\mathbf{x}}  
\newcommand{\bfy} {\mathbf{y}}  

\newcommand{\bv}{\begin{vugraph}}
\newcommand{\ev}{\end{vugraph}}
\newcommand{\bi}{\begin{itemize}}
\newcommand{\ei}{\end{itemize}}
\newcommand{\ben}{\begin{enumerate}}
\newcommand{\een}{\end{enumerate}}

\newcommand{\bean}{\begin{eqnarray*} }
\newcommand{\eean}{\end{eqnarray*} }
\newcommand{\bea}{\begin{eqnarray} }
\newcommand{\eea}{\end{eqnarray} }

\newcommand{\ba}{\begin{array} }
\newcommand{\ea}{\end{array} }

\newcommand{\bfe}{\mathbf{e}}

\newcommand{\bfs}{\mathbf{s}}
\newcommand{\bfv}{\mathbf{v}}

\newcommand{\bfr}{\mathbf{r}}
\newcommand{\bfm}{\mathbf{m}}
\newcommand{\bfc}{\mathbf{c}}
\newcommand{\bfz}{\mathbf{z}}

\newcommand{\calC}{\mathcal{C}}

\newcommand{\calL}{\mathcal{L}}
\newcommand{\calX}{\mathcal{X}}
\newcommand{\calY}{\mathcal{Y}}

\begin{document}

\title{Algebraic codes for Slepian-Wolf code design}

\author{
\authorblockN{Shizheng Li and Aditya Ramamoorthy}\\
\authorblockA{Department of Electrical and Computer Engineering,
Iowa State University,
Ames, Iowa 50011\\
Email: \{szli, adityar\}@iastate.edu}\thanks{This work was
supported in part by NSF grant CCF-1018148.} }

\maketitle

\begin{abstract}
Practical constructions of lossless distributed source codes (for
the Slepian-Wolf problem) have been the subject of much
investigation in the past decade. In particular, near-capacity
achieving code designs based on LDPC codes have been presented for
the case of two binary sources, with a binary-symmetric
correlation. However, constructing practical codes for the case of
non-binary sources with arbitrary correlation remains by and large open. From a practical perspective it is also interesting to consider coding schemes whose performance remains robust to uncertainties in the joint distribution of the sources.

In this work we propose the usage of Reed-Solomon (RS) codes for
the asymmetric version of this problem. We show that algebraic
soft-decision decoding of RS codes can be used effectively under
certain correlation structures. In addition, RS codes offer
natural rate adaptivity and performance that remains constant
across a family of correlation structures with the same
conditional entropy. The performance of RS codes is compared with
dedicated and rate adaptive multistage LDPC codes (Varodayan et
al. '06), where each LDPC code is used to compress the individual
bit planes. Our simulations show that in classical Slepian-Wolf
scenario, RS codes outperform both dedicated and rate-adaptive
LDPC codes under $q$-ary symmetric correlation, and are better
than rate-adaptive LDPC codes in the case of sparse correlation
models, where the conditional distribution of the sources has only
a few dominant entries. In a feedback scenario, the performance of
RS codes is comparable with both designs of LDPC codes. Our
simulations also demonstrate that the performance of RS codes  in
the presence of inaccuracies in the joint distribution of the
sources is much better as compared to multistage LDPC codes.
\end{abstract}

\vspace{-2mm}
\section{Introduction}
We consider the problem of practical code design for the
Slepian-Wolf problem.
Following the work of \cite{wynersol} that established the
equivalence between the Slepian-Wolf problem and channel coding, a
lot of research work has addressed this problem (see
\cite{xiongspmag} and its references). However, by and large most
of the work considers the case of two binary sources that are
related by an additive error. In this paper, we propose a coding
scheme for nonbinary sources using Reed-Solomon codes that works
under more general correlation models than an additive error
model. One previously proposed approach for compressing two
nonbinary sources is to use several LDPC codes, each for a bit
level of the binary image \cite{YangSXZ09} along with multistage
decoding. It requires the knowledge of the joint distribution and
the conditional distributions of the binary sources that
corresponding to the bit levels. It also requires the design of
multiple LDPC codes, multiple LDPC decodings at the terminal and
may suffer from error propagation. The multistage LDPC approach
breaks down the symbol level correlation to bit level
correlations. When the correlation is essentially at the symbol
level, multistage LDPC may not be the most suitable approach. In
this paper we evaluate the performance of RS codes and multistage
LDPC codes. We note that very few simulation results of multistage
LDPC codes for Slepian-Wolf problem on large alphabet sizes have
appeared in previous work. Turbo code-based design of nonbinary
SWC was proposed in \cite{ZhaoGarcia02} but only field size of
eight was considered. The work of \cite{MM10} proposed algebraic
codes for SWC using list decoding. Our algorithm uses soft
decoding and has better performance. In addition, we provide
simulations and comparisons with multistage LDPC codes.

In this paper, two scenarios are considered in our simulation. One
is the classical Slepian-Wolf scenario, where there is no feedback
from the decoder to the encoder. In the other scenario, there is
 feedback from the decoder to the encoder that tells the
encoder whether the decoding is successful. If the decoding fails,
the encoder will send more syndrome symbols. In this paper, we
consider two designs of multistage LDPC codes \cite{YangSXZ09},
(i) Dedicated codes for each bit source. The degree distributions
of the codes are optimized for AWGN channels and the codes are
generated by PEG algorithm \cite{PEG}. These codes do not offer
rate adaptivity. (ii) The rate adaptive codes designed in
\cite{VaroAG06}. The rate adaptivity in Slepian-Wolf problem
requires us to adapt the transmission rate by adapting the
syndrome length, rather than code length. If a low transmission
rate is not enough to decode the source, more syndrome symbols are
transmitted to the decoder and together with previously received
syndrome, the decoder attempts to decode. RS codes offer natural
rate-adaptivity by definition. Rate adaptive codes are useful in
the feedback scenario. Our simulations show that in the classical
Slepian-Wolf coding scenario, under $q$-ary symmetric correlation
models, RS codes outperform both designs of multistage LDPC codes.
Under sparse correlation models, RS codes perform better than rate
adaptive LDPC codes when the correlation resembles $q$-ary
symmetric models. In the feedback scenario, the performance of
rate-adaptive LDPC codes and RS codes are comparable under $q$-ary
symmetric channels but under sparse correlation model,
rate-adaptive LDPC codes perform better than RS codes. Moreover,
when the correlation given to the decoder is slightly different
from the true correlation model, RS codes suffer little but
multistage LDPC codes suffer significantly.


 This paper is
organized as follows. The preliminaries about RS codes and the
Koetter-Vardy decoding algorithm \cite{KV} are given in Section
\ref{sec:pre}. The RS code-based asymmetric SWC schemes are
described in Section \ref{sec:asym} and the performance
comparisons with a single LDPC codes are presented in Section
\ref{sec:compLDPC}. In Section \ref{sec:compMultiSW} and Section
\ref{sec:compMultiRA} the performance comparisons of RS codes and
multistage LDPC codes under two scenarios are presented
respectively. Section \ref{sec:conlu} concludes the paper.
\vspace{-2mm}
\section{Preliminaries}\label{sec:pre}
Let $F_q$ be a finite field and $q$ be a power of two. A $(n,k)$
RS code can be defined by its parity check matrix
$H_{ij} = (\alpha^i)^{j-1}, i = 1,\ldots,n-k,j=1,\ldots,n$,
where $\alpha$ is
a primitive element of $F_q$ and $n=q-1$. The code $\calC_{RS} =
\{\bfc\in F_q^n: H_{RS} \bfc = 0\}$.
Suppose $H_1, H_2$ are the parity check matrices of two RS codes
with rates $k_1/n, k_2/n$ respectively and $k_1\geq k_2$, by
definition, $H_1$ is a submatrix of $H_2$. As we shall see later,
this allows rate adaptivity for distributed source coding. An
equivalent definition of an RS code is given in terms of
polynomial evaluation. Given a message vector $\bfm$ of length
$k$, the encoded codeword is obtained by evaluating the message
polynomial $f_{\bfm}(\calX)$ (of degree $k-1$) at $n$ points
$\{1,\alpha,\alpha^2,\ldots,\alpha^{n-1}\}$. One only needs to specify
the code parameters $n$ and $k$ when designing codes.

Consider a channel coding scenario. A codeword $\bfc\in
\calC_{RS}$ is transmitted and the channel output is $\bfr$. Let
$\gamma_1,\ldots,\gamma_q$ be a fixed ordering of the elements
from $F_q$. The receiver computes the $q$-by-$n$ reliability
matrix $\Pi = \{\pi_{ij} = P(c_j = \gamma_i|r_j)\}$ based on the
information from the channel. The Koetter-Vardy soft decoding
algorithm \cite{KV} first computes a multiplicity matrix $M$ from
$\Pi$. The simplest choice is $M = \lfloor \lambda \Pi\rfloor$,
where $\lambda$ is a positive real number. Next, it constructs a
bivariate polynomial $Q_M(\calX,\calY)$ with minimal weighted
degree that passes through every point $(\alpha^{j-1},\gamma_i)$,
$m_{ij}$ times. These algebraic constraints can be given by $C(M)$
linear constraints, where $C(M) = \frac{1}{2} \sum_{i=1}^q
\sum_{j=1}^q m_{ij}(m_{ij}+1)$ is called the cost of $M$. Finally
it identifies all the factors of $Q_M(\calX,\calY)$ of type
$\calY-f(\calX)$, where $f(\calX)$ has degree no more than $k-1$.
Among these, it picks the candidate with the highest likelihood
based on the channel pmf. Note that the row index of $M$ can also
be given by an element from the $F_q$, i.e., $m_j(\beta) = m_{ij}$
if $\beta = \gamma_i$. The score of a vector $\bfv$ with respect
to a multiplicity matrix $M$ is defined to be
$S_M(\bfv)=\sum_{j=1}^n m_j(v_j)$. If the entries in $M$
corresponding to the transmitted codeword $\bfc$ have large
values, then $\bfc$ has high score w.r.t. $M$. It has been shown
\cite{KV} that as long as the score of a codeword $S_M(\bfc)\geq
\Delta_{1,k-1}(C(M))$, $\bfc$ will appear on the candidate list,
i.e., the decoding is successful. $\Delta_{1,k-1}(C(M))$ is
defined in \cite{KV} and depends on $k$ and $C(M)$ (increases with
them). \vspace{-2mm}
\section{RS codes for asymmetric SWC}\label{sec:asym}
Consider an asymmetric SWC scenario where source $X$ is available
at the terminal. If an RS code is used, the encoding for $\bfy$ is
its syndrome $\bfs = H\bfy$. The decoder needs to find the most
probable $\hat{\bfy}$ that belongs to the coset with syndrome
$\bfs$. Upon obtaining $\bfx$, the decoder finds the reliability
matrix $\Pi = \{\pi_{ij} = P(Y_j = \gamma_i|X_j = x_j)\}$ based on
the joint distribution. Then, use the multiplicity algorithms to
find a multiplicity matrix $M$. The simplest choice is $M =
\lfloor \lambda \Pi \rfloor$. If the RS code is powerful enough to
correct the errors introduced by the correlation channel, the
score $S_M(\bfy)$ should satisfy the score condition. We want to
obtain $\bfy$ from the matrix $M$ by interpolation and
factorization. Note that $\bfy$ is not a codeword but belongs to a
coset with syndrome $\bfs$. This requires us to modify the KV
algorithm appropriately. An approach to modify Guruswami and
Sudan's hard decision decoding algorithm \cite{GS99} to syndrome
decoding was proposed in \cite{LiRCommLet2010} and \cite{MM10}
independently. Our approach is motivated by them. Find a $\bfz$
belonging to the coset with syndrome $\bfs$. This can be done by
letting any $k$ entries in $\bfz$ to be zero and solve $H \bfz =
\bfs$. The uniqueness of the solution is guaranteed by the MDS
property of the RS code. Construct a shifted multiplicity matrix
$M'$ from $M$ according to $\bfz$, where $m'_j(\gamma_i) =
m_j(\gamma_i + z_j)$, or, equivalently, $m'_j(\gamma_i + z_j) =
m_j(\gamma_i)$, for $1\leq i\leq q, 1\leq j\leq n$. Interpolate
the $Q_{M'}(\calX,\calY)$ according to $M'$ as in KV algorithm and
find the list of candidate codewords $\calL_{\bfc}$ by
factorization. Adding $\bfz$ to each candidate codeword we obtain
the set of candidates $\calL_{\bfy}$ for $\bfy$.

\noindent {\it Claim}: $\bfy\in \calL_{\bfy}$ if $H\bfy = \bfs$ and $S_M(\bfy)\geq
\Delta_{1,k-1}(C(M))$.

\noindent {\it Proof}: The interpolation and factorization ensure that if a
codeword $\bfc$ is such that $S_{M'}(\bfc)\geq
\Delta_{1,k-1}(C(M'))$, $\bfc\in \calL_{\bfc}$. Note that each
column of $M'$ is just a permutation of the corresponding column
of $M$, so $C(M) = C(M')$ and $\Delta_{1,k-1}(C(M')) =
\Delta_{1,k-1}(C(M))$. If a vector $\bfy$ satisfies $H\bfy = \bfs$
and $S_M(\bfy)\geq \Delta_{1,k-1}(C(M))$, $\bfy+\bfz$ is a
codeword and $S_{M'}(\bfy+\bfz) = \sum_{j=1}^n
m'_j(y_j+z_j)=\sum_{j=1}^n m_j(y_j) =  S_M(\bfy)\geq
\Delta_{1,k-1}(C(M'))$, thus $\bfy+\bfz \in \calL_{\bfc}$. So
$\bfy\in \calL_{\bfy}$. 
\endproof

Next, the decoder performs ML decoding on $\calL_{\bfy}$ based on
$\Pi$. It is shown in the simulations that this step is almost
always correct. Thus, if $\bfy$ satisfies the score condition, the
decoding is successful (with very high probability). The
performance of the algorithm depends on the multiplicity
assignment, during which the correlation between the sources is
exploited.

\textbf{Remark}: 1) The soft information we used is the
conditional pdf $P(Y|X)$. It does not require the correlation
model to be additive. So it is suitable for more general
correlation models.

2) RS codes enable rate adaptivity easily because of the structure
of the parity check matrix. Suppose a syndrome $H_1\bfy$ is
available at the decoder but the decoding fails. The terminal
wants to know $H_2\bfy$, where $H_2$ has $(n-k_2)$ rows and
$k_1\geq k_2$. We can transmit additional inner products of $\bfy$
and newly added rows in $H_2$ and together with the syndrome
received previously, the decoder obtains the syndrome $H_2\bfy$.
Then the decoder works for a code with lower code rate.

\vspace{-2mm}
\section{Comparison with a single LDPC code}\label{sec:compLDPC} RS codes are Maximum Distance Separable (MDS) codes.
However, it is well known that RS codes are not
capacity-achieving over probabilistic channels such
as the BSC and the $q$-ary symmetric channel. On the other hand, LDPC codes
are capacity-achieving under binary symmetric channels. It is
expected and observed in simulation that for \textit{binary}
correlated sources, LDPC codes have better performance. However,
we expect that RS codes could be a better fit for sources over
large alphabets, at least for the channels that resemble
deterministic channels, e.g., $q$-ary symmetric channels.

One simple way to use LDPC codes in nonbinary Slepian-Wolf coding
is to use a single LDPC code to encode the binary image of the
nonbinary symbols. Consider a correlation model for sources $X$
and $Y$ expressed as $X = Y+E$, where $X,Y,E\in F_{512}$ such that
$E$ is independent of $X$ and the agreement probability $P_a =
P(E=0) = 1-p_e, P(E=\gamma)=p_e/(q-1)$ for nonzero $\gamma\in
F_{512}$. $X$ and $Y$ are uniformly distributed. This is called
$q$-ary symmetric correlation model. RS codes are defined over
$F_{512}$ with length 511. The LDPC codes for comparison have
length 4599 and a maximum variable node degree of 30 and were
generated using the PEG algorithm \cite{PEG}. For a given source
pair, we use one LDPC code and encode for the binary image of the
source outputs and the initial bit level LLR for belief
propagation decoding is found by appropriate marginalization. We
used three different code rates.
 For each code, we increase $P_a$ (decrease $H(Y|X)$)
until the frame error rate was less than $10^{-3}$ and recorded
the corresponding $H(Y|X)$ as the maximum $H(Y|X)$ that allows us to perform near error-free compression. The results are available
in Table \ref{tab:compLDPCRS}.
\begin{table}[h]
\vspace{-2mm} \centering \caption{\label{tab:compLDPCRS}
Comparison of RS codes and LDPC codes}
\begin{tabular}{|c|c|c|c|}
\hline $k/n$ & Tx Rate (bits/sym) & RS max $H(Y|X)$ & LDPC max $H(Y|X)$\\
\hline 0.2 & 7.2 & 5.3175 & 3.7855\\
\hline 0.3 & 6.3 & 4.3770 & 3.3740 \\
\hline 0.5 & 4.5 & 2.8474 & 1.7271 \\
\hline
\end{tabular}\vspace{-3mm}
\end{table}
We observe that LDPC has larger gap between the $H(Y|X)$ and the
actual transmission rate than RS codes. As expected, RS codes also
have a gap to the optimal rate. We also run the unique decoding
algorithm for RS codes (Berlekamp-Massey algorithm) and observe
that the performance is better than LDPC codes but worse than KVA.
\vspace{-3mm}
\section{Comparison with multistage LDPC
codes: Classial Slepian-Wolf scenario}\label{sec:compMultiSW}
\vspace{-1mm}
\subsection{Multistage LDPC codes}
Multistage LDPC codes have been proposed for
Slepian-Wolf coding for nonbinary alphabets in prior work
\cite{YangSXZ09}. To compress a source with alphabet size $q$, we
can view it as $r=\log_2 q$ binary sources. Suppose $X$ is known
at the terminal and the source $Y$ is represented as bit sources
$Y_{b_1}, Y_{b_2},\ldots, Y_{b_r}$.
The source transmits the syndromes of each bit source sequence,
$\bfs_k = H_k\bfy_{b_k} , k=1,2,\ldots,r$, where $H_k$ is the
parity check matrix of a LDPC code. At the decoder, the side
information $X$ is given, and to decode the $k$th bit source, the
previous decoded bit sources can also be used as side information,
based on which the initial LLR is computed.
The decoding requires us to
decode $r$ LDPC codes.

The design of optimized LDPC codes for our problem requires us to consider the individual bit level channels and the distribution of the input LLRs at each bit level. This is a somewhat complicated task and is part of ongoing work. Here we use the following two designs for
comparison.

\subsubsection{Dedicated LDPC codes}
We optimize the degree distribution using density evolution for
AWGN channel\footnote{As explained before, ideally we should run
density evolution for the actual bit level channel broken down
from the symbol level correlation channel. This is part of ongoing
work. In addition, we require a large number of codes in order to
match the required rates at the different bit levels. Since AWGN
optimized LDPC codes are known to have very good performance in
related channels such as the BSC, we chose to work with them for
the comparison.}. Then, the code of length 512 is designed by PEG
algorithm\footnote{We need to choose a block length for each LDPC
code so  that the comparison with the RS code of length 255 (8-bit
symbols) is fair. We chose a length of 512, that is approximately
$2\times 255$. With higher LDPC block lengths, one can expect
better performance.}. We design LDPC codes with rates
$0.02,0.04,0.06,\ldots, 0.90$, a total of 45 codes. These codes
are designed separately and do not provide rate adaptivity.

\subsubsection{Rate-adaptive LDPC codes}
Designed in \cite{VaroAG06}, these
irregular LDPC codes have length 6336 and the code
rate can be chosen among $\{0/66,1/66, \ldots, 64/66\}$. The
structure of their parity check matrices allow us to
 use them in a rate-adaptive
manner.  Note that these codes have a very high block length.

\vspace{-2mm}
\subsection{Simulation Setting}
We consider classical SWC scenario. Given a correlation model, we
gradually increase the transmission rate until the frame error
rate is less than $10^{-3}$. The decoder attempts decoding only
once. For LDPC codes, a frame is in error if one of the decodings
is in error. When we adjust the transmission rate, we adjust the
rate of the LDPC codes for each bit source, so that the FER for
each bit source are of the same order. To get the FER$<10^{-3}$ at
nonbinary symbol level, the FERs at the bit level are roughly
$10^{-4}$. For each rate configuration, we simulate until the
number of error frame is at least 100. The maximum iteration time
of the belief propagation algorithm is 100.
For RS codes, the field size $q = 256$ and the length $n = 255$.
$\lambda = 100.99$ in the multiplicity assignment. We increase the
transmission rate until the FER $<10^{-3}$. The decoder attempts
decoding only once.

\vspace{-2mm}
\subsection{$q$-ary symmetric correlation model}
The simulation results for $q$-ary ($q = 256$) symmetric
correlation model under different agreement probabilities are
given in Fig. \ref{fig:QaryData}(solid lines). The gaps between
actual transmission rates and $H(Y|X)$ are presented. Larger gap
indicates worse performance. We observe that under $q$-ary
symmetric correlation models RS codes outperform both types of
LDPC codes. This coincides with our intuition since the $q$-ary
symmetric is favorable for RS codes. Note that RS codes performs
better when the agreement probability $P_a$ is very high or very
low. For low $P_a$, a RS code with low rate is used and it is
observed before \cite{KV} that the Koetter-Vardy algorithm
performs better for low rate codes. When $P_a$ is very low, for
multistage LDPC codes, only a portion of bit sources can be
compressed, several bit sources need to be transmitted at rate
one.
\begin{small}
\begin{figure}[h]
\vspace{-3mm} \centering
\includegraphics[width=80mm,
clip=true]{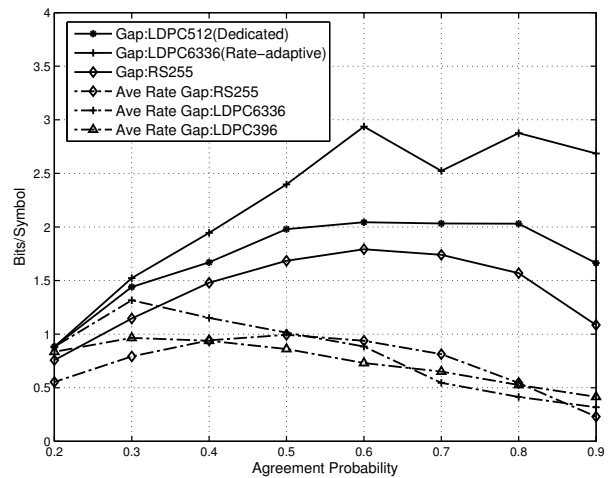} \caption{\label{fig:QaryData} The gap
between the transmission rate and $H(Y|X)$ for multistage LDPC and
RS codes under $q$-ary symmetric models. Solid line represents
classical SWC scenario and the dash-dot line represents feedback
scenario.} \vspace{-6mm}
\end{figure}
\end{small}
\vspace{-2mm}
\subsection{Sparse correlation model}
When the correlation model becomes more general, RS codes do not
always outperform LDPC codes. Under the correlation model where
each column of the conditional probability matrix $P(Y |X=j)$
contains a few dominant terms, it is possible that RS codes still
perform well. We call such kind of correlation models to be
sparse. We shall compare the performance of multistage LDPC codes
and RS codes under sparse correlation models defined as follows.

\begin{definition}
We say a conditional pdf $P(Y|X)$ is $(S,\epsilon)$-sparse if for
every $j = 1,\ldots,q$, $P(Y=i|X=j), i = 1, \ldots,q$ have $S$
entries that are greater than $\epsilon$.
\end{definition}

We are mostly interested in $(S,\epsilon)$-sparse conditional pdf
$P(Y|X)$ with $S\ll q$ and $\epsilon \ll 1$, i.e., for each $j$,
$P(Y=i|X=j)$ has few dominant entries. For those entries with
probability mass less than $\epsilon$, we assume that the
probabilities are the same.
When $X$ is uniformly distributed, the joint pdf is also sparse
and we call such a correlation model, a sparse correlation model.
For a $(S,\epsilon)$-sparse conditional pdf $P(Y=i|X=j)$, denote
the vector of the $S$ dominant entries by $D(j)$. We assume that
the dominant entries are the same for all $j$ and denote them by
$D$. For example, for a $q$-ary symmetric correlation model with
$q = 256$ and $P_a = 0.8$, $D = [0.8]$ and it is
$(1,10^{-3})$-sparse. For a fixed $D$, there are a lot of choices
of the locations of the dominant entries. We consider the
following dominant entry patterns.
%

The dominant entries can be put in the {\it diagonal form}, a
generalization of $q$-ary symmetric correlation model. The largest
entries are on the diagonal of the conditional pdf matrix and
other entries are put around them. For example, consider a joint
pdf with $(3,10^{-3})$-sparse conditional distribution and $D=
[0.1~0.6~0.1]$. When it is placed in the diagonal form,
$P(Y=j|X=j) = 0.6$ for all $j$, $P(Y=j-1|X=j) = 0.1$ for all $j$
except $j = 1$, $P(Y = j+1|X=j) = 0.1$ for all $j$ except $j =
256$ and $P(Y = 256|X=1) = P(Y=1|X=256) = 0.1$. All other entries
are $(1-0.1-0.6-0.1)/253 < 10^{-3}$.
The dominant entries in a conditional pdf is said to be in the
{\it random form} if $D$ is uniformly randomly placed in the
column $P(Y|X=j)$. Note that this
 randomness only appear in the determination of the pdf and it will be fixed during all transmissions. This
 correlation model is a model $Y = X + E$ where $E$ depends on
 $X$ (data dependent model).
Note that different placements of probability masses in the
columns of conditional distribution do not change the conditional
entropy $H(Y|X)$, and do not affect the performance of KV
algorithm for RS codes. But the performance of multistage LDPC
codes changes when the placement of probability masses changes. In
simulations, multistage LDPC codes performs better under diagonal
form conditional distribution than the random form.

Note that a dominant entry vector could have a number of forms. It
is hard to parameterize it using simple parameters. In our
simulations, we fix the length of $D$ to be three and there is one
distinguished large value in the vector. The vectors of dominant
entries in conditional pdf are presented in Table
\ref{tab:PeakAll}. They are the same for different $j$ in
$P(Y|X=j)$. Other than dominant entries, other entries have the
same probability. They are all $(3,0.0015)$-sparse conditional
pdfs. Source $X$ is uniformly distributed. For a vector of
dominant entries, we define peak factor to be the ratio between the maximum entry and the minimum
entry in the vector.
\begin{table}[h]\vspace{-3mm}
\begin{center}
\caption{\label{tab:PeakAll} The $D$ vectors used in the simulations.
}
\begin{tabular}{|c|c|c|c|}
\hline $D$ & PF & D & PF \\
\hline [0.15~ 0.6 ~0.15]& 4&   [0.1~ 0.6 ~0.1] &  6 \\
\hline [0.1~ 0.7 ~0.1] &  7 &[0.1~ 0.75 ~0.1] &  7.5 \\
\hline [0.1~ 0.79  ~0.1] & 7.9& [0.05~ 0.6 ~0.05] &  12 \\
\hline [0.05~ 0.7 ~0.05] &  14& [0.03~ 0.6 ~0.03]&  20\\
\hline
\end{tabular}
\end{center}\vspace{-5mm}
\end{table}
%

\begin{figure}[h]
\vspace{-1mm}
\begin{center}
\includegraphics[width=80mm,
clip=true]{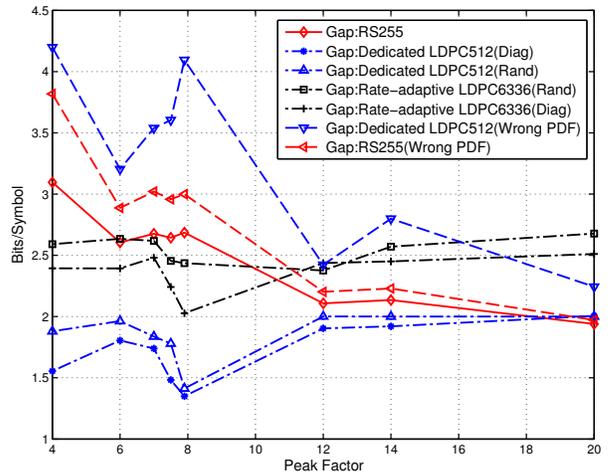} \caption{\label{fig:PeakPlot} The gap
between the actual transmission rate and the conditional entropy
for multistage LDPC codes and RS codes under sparse correlation
models. For RS codes, the performance under diagonal form
conditional distribution and random form conditional distribution
are the same.}
\end{center}\vspace{-5mm}
\end{figure}

We show our simulation results in Fig. \ref{fig:PeakPlot}, in an
ascending order of peak factor (PF). The plots do not look as
smooth as Fig. \ref{fig:QaryData}. This is because peak factor is
not a single parameter for the pdfs, e.g., for a fixed PF, there
could be multiple choices of the pdf and we choose one of them in
our simulation. The gaps between actual transmission rates and the
conditional entropies are presented. The alphabet size $q=256$.
Both random form and diagonal form conditional pdf are
investigated. For RS codes, the performance is the same under
these two forms.  We observe the following. The performance of RS
codes improves with the increase of the PF. RS codes perform
better than rate-adaptive LDPC codes under the correlation models
with large PF, while rate-adaptive LDPC codes perform better than
RS codes under the correlation models with small PF. However,
dedicated LDPC codes outperform RS for most of PF values.

We also investigate the situation where the decoder is given a
slightly different joint pdf. The actual pdf is in the diagonal
form. The pdf provided to the decoder has right locations for the
largest dominant entries but wrong (somewhat arbitrary) locations
for another two smaller dominant entries in $D$. In this case, the
performance of LDPC codes suffer a lot and RS codes suffer only a
little. The results are also presented in Fig. \ref{fig:PeakPlot}.
It is important to note that in this situation, RS codes in fact
perform better than multistage LDPC codes. In a practical setting
there may be situations where there are modeling errors or
incomplete knowledge about the joint pdf of the sources. Our
results indicate that RS codes are much more resilient to
inaccuracies in correlation models.

\section{Comparison with multistage LDPC codes: Feedback
scenario}\label{sec:compMultiRA}
\subsection{Simulation setting}
We consider the second scenario where the decoder feeds back some
information and the actual transmission rates are adapted such
that the decoder is able to decode. RS codes offer natural
rate-adaptivity and we compare their performance with the rate
adaptive LDPC codes designed in \cite{VaroAG06}. For multistage
LDPC codes, after receiving the binary syndromes from the encoder,
the decoder tries to decode from the first bit source. If it
fails, it requests more bits from the source and tries to decode
again. The decoder repeats this procedure until the first bit
source is decoded and then moves on to the second bit source and
works in a similar manner. It is guaranteed that the previously
decoded bits are always correct.
Two rate-adaptive LDPC codes are used, with length 6336 and 396,
both designed in \cite{VaroAG06}. For RS codes, if the decoder
fails (there is no codeword on the candidate list), it requests
more symbols from the source and tries again. The decoder repeats
this until the source sequence is decoded. The amount of feedback
is several bits per block for both LDPC codes and RS codes,
depending on the gap. But LDPC codes need more feedback since the
decoder needs to adjust rate for each bit source. We repeat this
experiment 500 times and record the minimum required transmission
rates. The simulation results are the average minimum required
rates and their standard deviation.\vspace{-2mm}
\subsection{$q$-ary symmetric correlation models} The gap of the average minimum transmission rate to
the conditional entropy is presented in Fig. \ref{fig:QaryData}
(dash-dot lines). RS outperform rate-adaptive LDPC codes when the
agreement probability is very high or very low. But for
intermediate $P_a$, multistage LDPC codes perform better. For LDPC
codes with length 6336, the standard deviations of the required
rates are in the range of 0.08 and 0.1, while LDPC codes with
length 396, the standard deviation are between 0.19 and 0.30. The
standard deviations of RS codes are between 0.13 and 0.32.

\vspace{-2mm}
\subsection{Sparse correlation models} 
\begin{figure}[h]
\vspace{-5mm}
\begin{center}
\includegraphics[width=78mm,
clip=true]{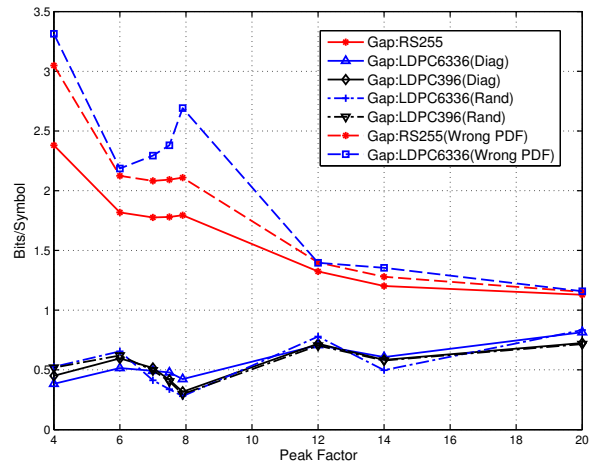} \caption{\label{Fig:AveRateRSSP} The
gap between the average minimum transmission rate and $H(Y|X)$ for
multistage LDPC and RS codes under sparse correlation models.}
\end{center}\vspace{-5mm}
\end{figure}

The gap of the average minimum transmission rate to
$H(Y|X)$ is presented in Fig. \ref{Fig:AveRateRSSP}. RS performs
worse than both multistage LDPC codes, although the performance
improves with the PF. The average rate performance is comparable
between LDPC codes with length 6336 and 396, and between diagonal
form and random form correlation models, but length 6336 codes are
much more stable, with standard deviation 0.06 to 0.1. RS codes
have standard deviation between 0.24 and 0.30, and length 396 LDPC
codes have standard deviation between 0.11 and 0.27.  The results
for the case where inaccurate pdfs are provided to the decoder are
also presented and we observe that RS codes are much more
resilient and perform better than LDPC codes with length 6336.

\vspace{-2mm}
\section{conclusion}\label{sec:conlu}
In this work we have proposed practical SW codes using RS codes.
Compared to multistage LDPC codes, RS codes are easy to design,
offer natural rate-adaptivity and allow for relatively fast
performance analysis. Simulations show that in classical SWC
scenario, RS codes perform better than both designs of multistage
LDPC codes under $q$-ary symmetric model and better than
rate-adaptive LDPC codes under the sparse correlation model with
high PF. In a feedback scenario, the performance of RS codes and
multistage LDPC codes are similar under $q$-ary symmetric model
but LDPC codes outperform RS codes under sparse correlation model.
An interesting conclusion is that RS codes are much more resilient
to inaccurate pdfs in both scenarios.

For symmetric Slepian-Wolf coding, if the correlation model is
given by additive error, i.e., $X = Y+E$, it is not hard to
propose a scheme that first recover the error vector $\bfe$ and
then recover the source sequences. The more interesting and
challenging problem is to apply algebraic approaches to more
general correlation models, where the problem can not be mapped to
a simple channel decoding problem. The problem remains open and
will be an interesting future work. \vspace{-2mm}
\bibliographystyle{IEEEtran}
\bibliography{AlgeDSCbib,DSC_new,RGBIB}
\end{document}